\documentclass[%
preprint,
amsmath,amssymb,
aps,
]{revtex4-2}

\usepackage{graphicx}
\usepackage{dcolumn}
\usepackage{bm}
\usepackage{hyperref}

\usepackage{diffcoeff, IEEEtrantools,
            amsmath, amssymb}

\DeclareMathOperator{\sech}{sech}

\makeatletter
\def\Dated@name{Accepted: }
\makeatother

\begin{document}

\preprint{APS/123-QED}

\title{Boundary-Bound Reactions: Pattern Formation\\with and without Hydrodynamics}

\author{Aiden Huffman}
\email{ahuffman@uwaterloo.ca}
\author{Henry Shum}%
\email{henry.shum@uwaterloo.ca}
\affiliation{%
 University of Waterloo,\\
 Waterloo, Canada
}%



\date{October 13th, 2023}


\begin{abstract}
We study chemical pattern formation in a fluid between two flat plates and the effect of such patterns on the formation of convective cells. This patterning is made possible by assuming the plates are chemically reactive or release reagents into the fluid, both of which we model as chemical fluxes. We consider this as a specific example of boundary-bound reactions. In the absence of coupling with fluid flow, we show that the two-reagent system with nonlinear reactions admits chemical instabilities equivalent to diffusion-driven Turing instabilities. In the other extreme, when chemical fluxes at the two bounding plates are constant, diffusion-driven instabilities do not occur but hydrodynamic phenomena analogous to Rayleigh--B\'enard convection are possible. Assuming we can influence the chemical fluxes along the domain and select suitable reaction systems, this presents a mechanism for the control of chemical and hydrodynamic instabilities and pattern formation. We study a generic class of models and find necessary conditions for a bifurcation to pattern formation. Afterwards, we present two examples derived from the Schnakenberg--Selkov reaction. Unlike the classical Rayleigh--B\'enard instability, which requires a sufficiently large unstable density gradient, a chemo-hydrodynamic instability based on Turing-style pattern formation can emerge from a state that is uniform in density. We also find parameter combinations that result in the formation of convective cells whether gravity acts upwards or downwards relative to the reactive plate. The wavenumber of the cells and the direction of the flow at regions of high/low concentration depend on the orientation, hence, different patterns can be elicited by simply inverting the device. More generally, our results suggest methods for controlling pattern formation and convection by tuning reaction parameters. As a consequence, we can drive and alter fluid flow in a chamber without mechanical pumps by influencing the chemical instabilities.
\end{abstract}

\maketitle


\section{\label{sec:introduction}Introduction}

Chemical pattern formation in reaction--diffusion systems has been of interest since Turing's original paper on chemical morphogenesis \cite{turing}. Turing demonstrated that nonlinear reactions with sufficiently asymmetric diffusion rates between chemical reagents can have unstable homogeneous steady states; concentration fields tend to approach steady, inhomogeneous distributions. His paper is a classical example of how breaking symmetry can cause a dissipative process such as diffusion to result in more order. The characteristics of a diffusion-driven instability are that the reaction--diffusion system possesses a homogeneous steady state that is linearly stable to small homogeneous perturbations of the concentration fields but unstable to spatially varying perturbations. For a two-reagent reaction--diffusion system given by
\begin{subequations}
    \begin{align}
        \diffp{Q}{t} &= f(Q,R) + \nabla^2 Q,\\
        \diffp{R}{t} &= g(Q,R) + D\nabla^2 R,
    \end{align}
\end{subequations}
for which there exist constants $Q^*, R^*$ satisfying $f(Q^*,R^*)=0=g(Q^*,R^*)$, the Turing conditions for pattern formation can be summarized as
\begin{IEEEeqnarray*}{rClCrCl}
    f_Q+g_R &<& 0, &\qquad& f_Qg_R-f_Rg_Q &>& 0,\\
    Df_Q+g_R &>& 0, &\qquad& f_Qg_R-f_Rg_Q &<& \frac{(Df_Q+g_R)^2}{4D},
\end{IEEEeqnarray*}
where subscripts denote partial derivatives evaluated at the homogeneous steady state $Q=Q^*,\, R=R^*$. Together, these conditions require $D\neq 1$ in order for pattern formation to occur. Conditions for Turing instabilities in systems of more than two reacting and diffusing components have also been determined~\cite{satnoianu_turing_2000}.

Recent extensions of Turing-style instabilities include active membrane models or bulk--surface reaction--diffusion systems \cite{membrane_bound_tp, Gomez2021, Madzvamuse2015}. Among many interesting properties, these works generally require smaller critical diffusion ratios for pattern formation. In particular, Levine et al. \cite{membrane_bound_tp} presented a model with a critical diffusion ratio of one. This result differs from the classical Turing instability, which generally requires large ratios of diffusion coefficients, suggesting that patterning may be easier to realize than the classical model predicts.

In early experiments that demonstrated chemical Turing patterns, reactions took place in a gel reactor~\cite{sustained-turing-type-pattern, lengyel_turing_1993}. One of the reasons for this was to suppress convection of the reagents, which could alter the stability of Turing patterns~\cite{convective_turing_patterns, convection_to_stability} or give rise to other chemo-hydrodynamic instabilities in more general settings~\cite{chemoconvection, taggart_bbr, de2012introduction, chemically-driven-convection, hydrodynamic-flows-traveling-with-chemical-waves, budroni, budroni2}.

While many chemical reactions take place in bulk fluid, it is also possible to restrict reactions to occur only at a solid boundary by immobilizing the necessary enzymes on the walls of the chamber. We use the term `boundary-bound' to refer to such reactions. A combination of theoretical and experimental work has shown that, through buoyancy effects, boundary-bound chemical reactions can produce fluid flow speeds on the order of micrometers per second in enclosed chambers with a fluid depth of around 1\,mm~\cite{self-powered-enzyme-micropumps, convective-flow-reversal, solutal-and-thermal-buoyancy-effects-in-self-powered-micropumps}. A benefit of using boundary-bound reactions as opposed to bulk reactions is that enzymes can be immobilized in customizable patterns, allowing the generated flow fields to be controlled. It has been proposed and demonstrated that the fluid flow induced by immobilized enzymes and catalysts can be used for performing tasks such as transport of microparticles and assembly of complex microstructures~\cite{using-chemical-pumps-and-motors-to-design-flows, harnessing-catalytic-pumps-for-directional-delivery}. Moreover, having controlled time-dependent flow would allow effective mixing~\cite{optimal-control-of-mixing}, which is notoriously challenging in microfluidic devices. 

The boundary-bound reaction systems mentioned above relied on an imposed symmetry breaking, either by introducing reagents at one side of the chamber or by immobilizing enzyme on part of the bottom surface. This generates a horizontal density gradient, which leads to convective fluid flows. Fluid flows can also be achieved through chemo-hydrodynamic instabilities that spontaneously break horizontal symmetry in a manner similar to the classical Rayleigh--B\'enard and Rayleigh--Taylor instabilities. Bdzil and Frisch~\cite{bdzil_bbr} performed a linear stability analysis of a layer of fluid between horizontal plates, one of which catalyzes a dissociation reaction. It was shown that such systems exhibit instabilities that are formally equivalent to the Rayleigh--B\'enard problem in the limit of high reaction rates but have lower thresholds for instability at finite reaction rates. Subsequent theoretical studies also described oscillatory instabilities and convective cells in the same system~\cite{chemically-driven-convection, taggart_bbr}. A different reaction system exhibiting chemoconvection was studied experimentally and theoretically by Bees et al.~\cite{chemoconvection}.

Apart from oscillations arising from chemo-hydrodynamic instabilities, there are examples of oscillatory chemical reaction systems, such as the Belousov--Zhabotinsky reaction~\cite{bz-murray}. Shklyaev et al. showed that an oscillating system of boundary-bound chemical reactions can exhibit enhanced amplitudes of oscillation or altered frequencies when coupled to fluid advection via buoyancy effects \cite{enhancement_of_chem_oscillations}. The feedback between chemical reactions and fluid flow also leads to the potential to regulate reactions by imposed and self-generated flows; theoretical modelling has demonstrated that reaction-induced convection increases reaction rates~\cite{Manna2022} and oscillatory chemical systems can exhibit amplified oscillations in the presence of resonant oscillatory flows~\cite{resonant-amp-of-enzymatic-chem-osc}.

Other works studying chemo-hydrodynamics have also focused on oscillatory behaviour, traveling waves \cite{hydrodynamic-flows-traveling-with-chemical-waves}, or designs where the chemical reagents are initially separated with a shared fluid interface \cite{budroni, budroni2, chemo-hydrodynamic-patterns-and-instabilities}. The latter works have also demonstrated how simple linear reactions in the bulk can produce spatio-temporal patterns when a hydrodynamic effect is included. We direct interested readers to De Wit et al.~\cite{de2012introduction} for a review.  

In contrast, our present work considers a horizontally uniform system with boundary-bound reactions on one or two of the confining horizontal plates. We focus on diffusion-driven chemical instabilities as a mechanism for breaking symmetry and obtaining horizontal patterns. We also investigate chemo-hydrodynamic instabilities by combining this reaction system with buoyancy forces associated with the concentrations of chemical species. Parallels in this design can be drawn from classical fluid instabilities such as Rayleigh--B\'enard convection, the Rayleigh--Taylor instability, and the double diffusive instability \cite{radko2013double, kundu2002fluid}. Here, the chemical reagents play the same role as the temperature and salt gradients, which produce various convective instabilities. 

To begin, we derive the necessary conditions for the existence of a Turing-style diffusion-driven instability in a class of boundary-bound reactions; to our knowledge this is the first analysis of this kind for these models. By considering a reaction that admits a Turing instability we make it possible for the chemical instability to influence the fluid behaviour and present an alternative mechanism for the formation of convective cells. The analytical results are demonstrated numerically with a focus on mechanisms for controlling the pattern forming instability. When the hydrodynamic effect is included, we find that Turing-style instabilities are a sub-case of the buoyancy-driven instability and that control of the chemical instability translates to control of the formation of convective cells. We also present a system that is chemically stable in the absence of buoyancy forces, has no density variations in the marginally stable steady state, and still admits a buoyancy-driven instability past a critical Rayleigh number. This is neither a purely chemical instability nor a classical Rayleigh--B\'enard instability where a notion of a heavy top and light bottom in the background state exists. Finally, we find parameter combinations that result in the formation of convective cells whether gravity acts upwards or downwards relative to the reactive plate. The wavenumber of the cells and the direction of the flow at regions of high/low concentration depend on the orientation, hence, different patterns can be elicited by simply inverting the device. These latter two results highlight the unique characteristics of the formation of convective cells in chemo-hydrodynamics.

\section{\label{sec:gm}General Model}

\begin{figure}
    \centering
    \includegraphics[width=0.6\textwidth]{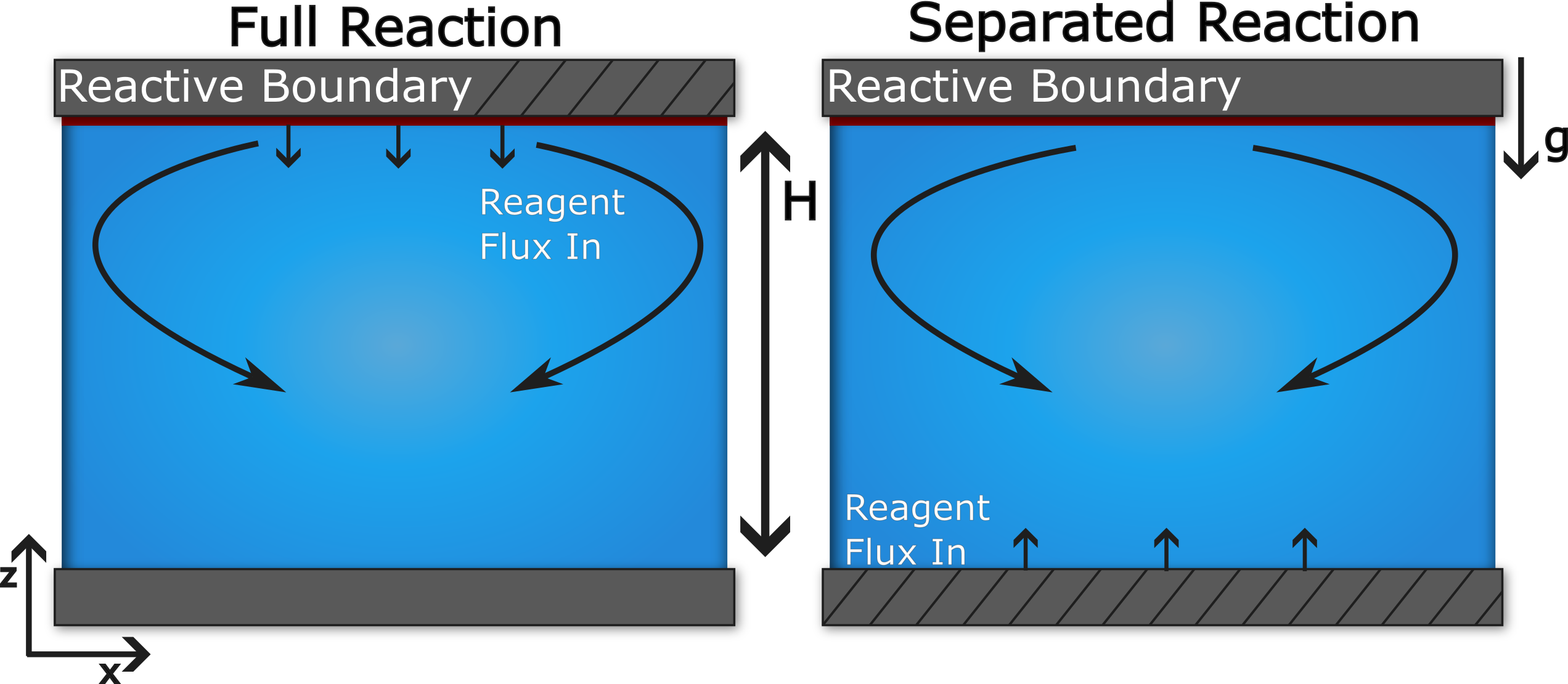}
    \caption{Schematic for the boundary-bound reaction models investigated in this work. In both cases, the reactions occur on the surface of the top plates. The separated reaction is derived from the full reaction by breaking off the constant fluxes and moving them to the lower plate.}
    \label{fig:schematic_bbr}
\end{figure}

We consider a domain consisting of a fluid between two flat plates. We will assume that these plates are separated by a vertical distance $H$ that is small relative to their horizontal length. As a consequence, we treat the domain as unbounded in the horizontal directions. The model formulation is identical in two and three spatial dimensions apart from the physical units of some parameters. A simple two-dimensional schematic is provided in Figure \ref{fig:schematic_bbr} for the specific models considered in this manuscript. We investigate two different models which we refer to as the full reaction and the separated reaction. To motivate these two classes of models, we note that it is reasonable to supply a constant, continuous source of reagents to the system if long-lived behaviour is desired. Experimentally, reagents could be introduced through a membrane or hydrogel, or as a reaction product from a precursor species that is assumed to remain in excess. In our models, we assume that reagents enter through one of the horizontal plates with a constant flux. There is also one horizontal plate that is catalytic where concentration-dependent reactions take place. The term `full reaction' refers to the model in which the constant fluxes occur at the same boundary as the other reactions. We use the term `separated reaction' when the constant fluxes are placed on the opposite boundary to the other reaction terms.

If we assume that density variations in the fluid are produced by the presence of two chemical reagents $(Q,R)$ and that these density variations are small, then the resulting fluid flows can be modelled using the Boussinesq approximation for the incompressible Navier--Stokes equations. We will assume that the reagents undergo exponential decay in the bulk with rate constants $(\eta_Q,\eta_R)$, diffuse with rates $(d_Q, d_R)$, and are advected by the fluid flow. Therefore, the bulk behaviour can be modelled via
\begin{subequations}\label{eq:reaction_eq}
    \begin{align}
        \frac{\mathrm{D}Q}{\mathrm{D}t} &= -\eta_Q Q + d_Q\nabla^2 Q,\\
        \frac{\mathrm{D}R}{\mathrm{D}t} &= -\eta_R R + d_R\nabla^2 R,\\
        \rho_0\frac{\mathrm{D}\mathbf{u}}{\mathrm{D}t} &= -\nabla p + \mu \nabla^2 \mathbf{u} + \rho_0 g (\beta_Q Q + \beta_R R)\mathbf{e}_z,\label{eq:momentum_eq}\\
        \nabla\cdot\mathbf{u} &= 0,
    \end{align}
\end{subequations}
where $\frac{\mathrm{D}}{\mathrm{D}t}=\frac{\partial}{\partial t} + \mathbf{u}\cdot \nabla$ denotes the convective derivative operator, $\rho_0$ is the fluid density in the absence of reagents, $p$ is the fluid pressure, $\mu$ is the dynamic viscosity of the fluid, $g$ is the signed acceleration due to gravity ($g<0$ for gravity acting downwards), and $\beta_Q$ and $\beta_R$ are the solutal expansion coefficients for $Q$ and $R$, respectively. At the boundary, the enzyme-driven reactions (or other processes that add or remove regents) are described by chemical fluxes. In combination with the no-slip condition for the fluid flow, the full set of boundary conditions is,
\begin{equation}
    \begin{IEEEeqnarraybox}[][c]{rClCrCl}
      -d_Q\diffp[1]Qz\bigg|_{z=0} &=& f_0(Q,R), &\qquad& d_Q\diffp[1]Qz\bigg|_{z=H} &=& f_1(Q,R),\\
      -d_R\diffp[1]Rz\bigg|_{z=0} &=& g_0(Q,R), &\qquad& d_R\diffp[1]Rz\bigg|_{z=H} &=& g_1(Q,R),\IEEEyesnumber\\
      \mathbf{u}\bigg|_{z=0} &=& \mathbf{0}, &\qquad& \mathbf{u}\bigg|_{z=H} &=& \mathbf{0}.
    \end{IEEEeqnarraybox}
\end{equation}

We non-dimensionalize the system using the scalings
\begin{equation}
   \begin{IEEEeqnarraybox}[][c]{rClCrClCrCl}
        \mathbf{x} &=& H\hat{\mathbf{x}}, &\qquad& t &=& \frac{H^2}{d_Q}\hat{t}, &\qquad& Q &=& Q_s\hat{Q},\\
        R &=& R_s\hat{R}, &\quad& \mathbf{u} &=& \frac{d_Q}{H}\hat{\mathbf{u}}, &\quad& p &=& \frac{d_Q^2\rho_0}{H^2}\hat{p},\\
    \end{IEEEeqnarraybox}
\end{equation}
where $Q_s$ and $R_s$ are arbitrary concentration scales that will be determined by reactions at the boundaries. Letting $\beta^{\max} = \max\{|\beta^Q Q_s|,|\beta^R R_s|\}$, the dimensionless parameters are:
\begin{equation}
   \begin{IEEEeqnarraybox}[][c]{rClCrClCrClCrCl}
        D &=& \frac{d_R}{d_Q}, &\qquad& \hat{\eta}_Q &=& \frac{H^2}{d_Q}{\eta}_Q, &\qquad& \hat{\eta}_R &=& \frac{H^2}{d_Q}{\eta}_R,\\
        \mathrm{Sc} &=& \frac{\mu}{\rho_0 d_Q}, &\qquad& \mathrm{Gr} &=& \frac{\rho_0^3 g\beta^{\max} H^3}{\mu^2}, &\qquad& \delta^Q &=& \frac{\beta^Q Q_s}{\beta^{\max}}, &\qquad& \delta^R &=& \frac{\beta^R R_s}{\beta^{\max}}.
    \end{IEEEeqnarraybox}
\end{equation}
Here, $\mathrm{Sc}$ is the Schmidt coefficient, $\mathrm{Gr}$ is the Grashof number and $D$ is the diffusion ratio between the two chemical reagents. By defining $\delta^Q$ and $\delta^R$ relative to the maximum modulus of the reagent expansion coefficients, we ensure that one of $\delta^Q$ or $\delta^R$ is equal to $\pm 1$ and the other is bounded in absolute value by $1$. In Section \ref{sec:w_hydrodynamics} we will refer to the case where $|\delta^Q|=1$ or $|\delta^R|=1$ as $Q$- or $R$-driven flow, respectively.

Applying our non-dimensionalization and dropping the hats on dimensionless variables, \eqref{eq:reaction_eq} becomes
\begin{subequations}\label{eq:full_bbr}
    \begin{align}
        \frac{\mathrm{D}Q}{\mathrm{D}t} &= -\eta_Q Q + \nabla^2 Q,\\
        \frac{\mathrm{D}R}{\mathrm{D}t} &= -\eta_R R + D\nabla^2 R,\\
        \frac{\mathrm{D}\mathbf{u}}{\mathrm{D}t} &= -\nabla p + \mathrm{Sc} \nabla^2 \mathbf{u} + \mathrm{Sc}^2 \mathrm{Gr}(\delta_Q Q + \delta_R R)\mathbf{e}_z,\\
        \nabla\cdot\mathbf{u} &= 0.
    \end{align}
\end{subequations}
The boundary conditions for the non-dimensionalized system are:
\begin{equation}\label{eq:general_bc_bbr}
    \begin{IEEEeqnarraybox}[][c]{rClCrCl}
      -\diffp[1]Qz\bigg|_{z=0} &=& \gamma f_0(Q,R), &\qquad& \diffp[1]Qz\bigg|_{z=1} &=& \gamma f_1(Q,R),\\
      -D\diffp[1]Rz\bigg|_{z=0} &=& \gamma g_0(Q,R), &\qquad& D\diffp[1]Rz\bigg|_{z=1} &=& \gamma g_1(Q,R),\IEEEyesnumber\\
      \mathbf{u}\bigg|_{z=0} &=& \mathbf{0}, &\qquad& \mathbf{u}\bigg|_{z=1} &=& \mathbf{0}.
    \end{IEEEeqnarraybox}
\end{equation}
Here, we view $\gamma$ as a general positive scaling for the chemical fluxes at the boundaries.

Suppose that the system has an unpatterned (i.e., the concentration and pressure fields depend only on the $z$ coordinate and not on horizontal coordinates) steady state without flow. We perform a linear stability analysis around this steady state considering a perturbation of the form,
\begin{equation}\label{eq:perturbation}
\begin{aligned}
&Q(\mathbf{x},t) = Q_0(z) + \epsilon Q_1(\mathbf{x},t), &&R = R_0(z) + \epsilon R_1(\mathbf{x},t),\\
&\mathbf{u}(\mathbf{x},t) = \epsilon \mathbf{u}_1(\mathbf{x},t), && p(\mathbf{x},t) = p_0(z) + \epsilon p_1(\mathbf{x},t).
\end{aligned}
\end{equation}
The resulting system can be simplified by eliminating the pressure term and dropping the horizontal dynamics. Details of this technique can be found in Appendix \ref{ap:A}. Considering the ansatz,
\begin{equation}
  \begin{bmatrix}
    Q_1(\mathbf{x},t)\\ R_1(\mathbf{x},t)\\ u_{1,z}(\mathbf{x},t)\\ p_1(\mathbf{x},t)
  \end{bmatrix} = \begin{bmatrix}
    q(z)\\ r(z)\\ w(z)\\ p(z)
  \end{bmatrix}e^{\lambda t + i(k_x x + k_y y)},
\end{equation}
where $u_{1,z}$ is the vertical component of $\mathbf{u}_1$, we obtain the perturbation problem:
\begin{subequations}\label{eq:perturbed_equations}
  \begin{align}
    \left(\diff[2]{}{z} - (\lambda + \eta_Q + k^2)\right) q &= w\diff{Q_0}{z},\\[15pt]
    \left(\diff[2]{}{z} - \left(\frac{\lambda + \eta_R}{D} + k^2\right)\right)r &= w\diff{R_0}{z},\\[15pt]
    \left[\left(\diff[2]{}{z} - k^2\right)^2 - \frac{\lambda}{\mathrm{Sc}}\left(\diff[2]{}{z}-k^2\right)\right] w &= k^2\mathrm{Ra}(\delta^Q q+\delta^R r),
  \end{align}
\end{subequations}
where $k^2 = k_x^2 + k_y^2$ and $\mathrm{Ra} = \mathrm{Gr}\cdot \mathrm{Sc}$ is the solutal Rayleigh number, subject to the boundary conditions,
\begin{equation}\label{eq:bbr_bcs}
  \begin{IEEEeqnarraybox}[][c]{rClCrCl}
      -\diff[1]qz\bigg|_{z=0} &=& \gamma (f_{0,Q}\,q+f_{0,R}\,r), &\qquad& \diff[1]qz\bigg|_{z=1} &=& \gamma (f_{1,Q}\,q+f_{1,R}\,r),\\
      -D\diff[1]rz\bigg|_{z=0} &=& \gamma (g_{0,Q}\,q+g_{0,R}\,r), &\qquad& D\diff[1]rz\bigg|_{z=1} &=& \gamma (g_{1,Q}\,q+g_{1,R}\,r),\IEEEyesnumber\\
      w\bigg|_{z=0,1} &=& 0, &\qquad& \diff[]wz\bigg|_{z=0,1} &=& 0.
    \end{IEEEeqnarraybox}
\end{equation}
In \eqref{eq:bbr_bcs} and below, the notation $h_{j,C}$ represents the partial derivative of the function $h_i$ with respect to $C$ evaluated at the unpatterned steady state concentrations $Q=Q_0(j), R=R_0(j)$, for all combinations of functions $h=f,g$, boundaries $j=0,1$, and concentration labels $C=Q,R$. For two spatial dimensions, we remove the $y$ variable and set $k_y=0$.

To begin, we assume that the Rayleigh number is small and that the density variations are negligible. This assumption reduces the problem to studying pattern formation without convection.

\section{\label{sec:pattern_formation}Pattern Formation}

\noindent
When it exists, the general solution to \eqref{eq:perturbed_equations} without flow is of the form
\begin{subequations}\label{eq:bbr_gen_sol}
\begin{align}
  q(z) &= a_1 \cosh(\Gamma_{1,\lambda} z) + a_2\sinh(\Gamma_{1,\lambda} z),\label{eq:bbr_gen_q_sol}\\
  r(z) &= b_1 \cosh(\Gamma_{2,\lambda} z) + b_2\sinh(\Gamma_{2,\lambda} z),\label{eq:bbr_gen_r_sol}\\
  w(z) &= 0,\label{eq:bbr_gen_w_sol}
\end{align}
\end{subequations}
where
\begin{subequations}\label{eq:Gamma_def}
  \begin{align}
  \Gamma_{1,\lambda} &= \sqrt{\lambda + \eta_Q + k^2},\\
  \Gamma_{2,\lambda} &= \sqrt{\frac{\lambda + \eta_R}{D} + k^2}.
  \end{align}
\end{subequations}
Below, we use the notation $\Gamma_i$ to indicate the values of $\Gamma_{i,\lambda}$ with $\lambda = 0$, for $i=1,2$. The boundary conditions~\eqref{eq:bbr_bcs} can be expressed as a homogeneous linear system of four equations in $a_i,\ b_i$. This system must be degenerate for non-trivial solutions to exist.

Consider the case where reactions for both $Q$ and $R$ occur at a shared (upper) boundary so that $f_1$ and $g_1$ have non-zero partial derivatives. At the bottom boundary, we will assume that $f_0$ and $g_0$ are constant (possibly zero), implying that their partial derivatives vanish. Given $\lambda$ and $k$, the resulting system of equations has a non-trivial solution if and only if
\begin{IEEEeqnarray}{rCl}
    0 &=& \gamma^2\det(J_1)\cosh(\Gamma_{1,\lambda})\cosh(\Gamma_{2,\lambda}) - \gamma(D\Gamma_{2,\lambda}\cosh(\Gamma_{1,\lambda})\sinh(\Gamma_{2,\lambda})f_{1,Q}\nonumber\\
    && \negmedspace{} + \Gamma_{1,\lambda}\sinh(\Gamma_{1,\lambda})\cosh(\Gamma_{2,\lambda})g_{1,R}) + D\Gamma_{1,\lambda}\Gamma_{2,\lambda}\sinh(\Gamma_{1,\lambda})\sinh(\Gamma_{2,\lambda}),\label{eq:shared_boundary}    
\end{IEEEeqnarray}
where $J_1 =\begin{bmatrix}
        f_{1,Q} & f_{1,R}\\
        g_{1,Q} & g_{1,R}
    \end{bmatrix}$.
    
Expressed in this form, we can show that $D\neq 1$ is necessary to make a finite wavenumber the most unstable mode. Suppose that there exist a (possibly) complex $\lambda$ and real $k > 0$ that satisfy~\eqref{eq:shared_boundary} with $\mathrm{Re}(\lambda)\geq 0$. Then, any other pair $(\lambda',k')$ such that $\Gamma_{i,\lambda}(k)=\Gamma_{i,\lambda'}(k')$ for $i=1,2$ would also satisfy~\eqref{eq:shared_boundary}. From~\eqref{eq:Gamma_def}, we note that defining $\lambda' = \lambda + k^2$ and $k'=0$ satisfies this requirement. Since $\mathrm{Re}(\lambda') > \mathrm{Re}(\lambda)$, the long wavelength instability $k'=0$ is more linearly unstable.

If we are concerned with transitions from a steady unpatterned state to a steady pattern with finite wavenumber, then we restrict ourselves to the case where $\lambda$ is real. Close to such a transition, we may assume that $\Gamma_{1,\lambda}$ and $\Gamma_{2,\lambda}$ are real and positive. Treating the entries of $J_1$ as independent parameters and assuming that $\det(J_1)\neq 0$, then \eqref{eq:shared_boundary} reduces to a quadratic equation for $\gamma$. Requiring that the global flux $\gamma$ be real, we obtain the condition that either $f_{1,R}$ has the same sign as $g_{1,Q}$ or
\begin{small}
\begin{equation}
  \left| f_{1,Q} - \frac{\Gamma_{1,\lambda}\tanh(\Gamma_{1,\lambda})}{D\Gamma_{2,\lambda}\tanh(\Gamma_{2,\lambda})}g_{1,R} \right| \geq 2\sqrt{-\frac{\Gamma_{1,\lambda}\tanh(\Gamma_{1,\lambda})}{D\Gamma_{2,\lambda}\tanh(\Gamma_{2,\lambda})}f_{1,R}\,g_{1,Q}}.
  \label{eq:sb_real_decay_condition}
\end{equation}
\end{small}

We seek marginally stable solutions, i.e., those with $\lambda = 0$. For the special case $\eta_R = 0$, \eqref{eq:shared_boundary} can be rearranged to obtain an explicit formula for the diffusion ratio when a given wavenumber $k>0$ is marginally stable,
\begin{equation}
    D_{c}(k) = \frac{\gamma \left(\Gamma_{1} g_{1,R} \tanh{\left(\Gamma_{1} \right)} - f_{1,Q} g_{1,R} \gamma + f_{1,R} g_{1,Q} \gamma\right)}{k\tanh{\left(k \right)} \left(\Gamma_{1} \tanh{\left(\Gamma_{1} \right)} - f_{1,Q} \gamma\right)},\label{eq:crit_D_sh}
\end{equation}
which we refer to as the critical diffusion ratio for the wavenumber $k$.

Assuming positive diffusion coefficients, we seek a positive value for $D_{c}(k)$, which requires that
\begin{equation}
    (\Gamma_1\tanh(\Gamma_1)-f_{1,Q}\gamma)(\Gamma_1g_{1,R}\tanh(\Gamma_1)-\gamma\text{det}(J_1)) > 0.\label{eq:d_conditions_sh}
\end{equation}
A critical diffusion cannot be found if this inequality is not satisfied for some $k$. 

In Section \ref{sec:exampels_wo_hydrodynamics}, we will present two examples of the case above with a shared boundary for reactions. If we consider an alternative setup where $g_{0}$ and $f_{1}$ have non-trivial partial derivatives while $g_1$ and $f_0$ are constant (non-constant reactions on opposite boundaries), then we would obtain the necessary condition for solutions with $\lambda\approx 0$ that either $f_{1,R}$ has the same sign as $g_{0,Q}$ or
\begin{equation}
    \bigg|f_{1,Q}-\frac{\Gamma_{1,\lambda}\tanh(\Gamma_{1,\lambda})}{D\Gamma_{2,\lambda}\tanh(\Gamma_{2,\lambda})}g_{0,R}\bigg| > 2\sqrt{-\frac{\Gamma_{1,\lambda}\tanh(\Gamma_{1,\lambda})}{D\Gamma_{2,\lambda}\cosh(\Gamma_{1,\lambda})\sinh(\Gamma_{2,\lambda})}f_{1,R}g_{0,Q}}\,,
\end{equation}
and under the same conditions we used in the derivation of $D_{c}(k)$ above, the critical diffusion ratio for this setup is positive only if
\begin{equation}
     (\Gamma_1\tanh(\Gamma_1)-f_{1,Q}\gamma)(\Gamma_1g_{0,R}\tanh(\Gamma_1)-\gamma\text{det}(J_2)) > 0,\label{eq:d_conditions_op}
\end{equation}
where $J_2 = \begin{bmatrix}
    f_{1,Q} & f_{1,R}\sech(\Gamma_1)\\
    g_{1,Q}\sech(k) & g_{1,R}
\end{bmatrix}.$

\subsection{Limiting Cases of Constant and Linear Reactions}

In the simple case of the reaction function $f_1$ being constant (independent of concentrations $Q$ and $R$), Equation \eqref{eq:shared_boundary} reduces to
 \begin{equation}
     \gamma \cosh(\Gamma_{2,\lambda})g_{1,R} - D\Gamma_{2,\lambda}\sinh(\Gamma_{2,\lambda}) = 0.
 \end{equation}
If this condition is satisfied for some $\lambda$ and finite wavenumber $k$, then it is also satisfied for other values $\lambda'$ and $k'$ if
\[\Gamma_{2,\lambda'} = \Gamma_{2,\lambda} = \sqrt{\frac{\lambda + \eta_R}{D} + k^2}\,;\] 
in particular, the wavenumber $k' = 0$ has a solution with $\lambda' = \lambda + Dk^2 > \lambda$. We conclude that no short wavelength instability will become unstable before a long wavelength one. A similar argument can be made in the case where $g_1$ is a constant function.

The next simplest classes of reaction functions $f_1, g_1$ are linear in reagent concentrations. While it is possible for Turing instabilities to occur in these cases, the linearly unstable eigenfunction mode would grow without bound, leading to unphysical behaviour. Linear reactions are, therefore, only valid models within restricted regimes; more complete models include nonlinear terms that prevent concentrations from becoming negative or growing without bound~\cite{mathematical-biology}.

With these considerations, we study a classical example of a nonlinear reaction system known to give rise to Turing instabilities in the following section.

\subsection{Schnakenberg--Selkov Examples}\label{sec:exampels_wo_hydrodynamics}

\begin{figure}[htpb]%
\centering
\includegraphics[width=\textwidth]{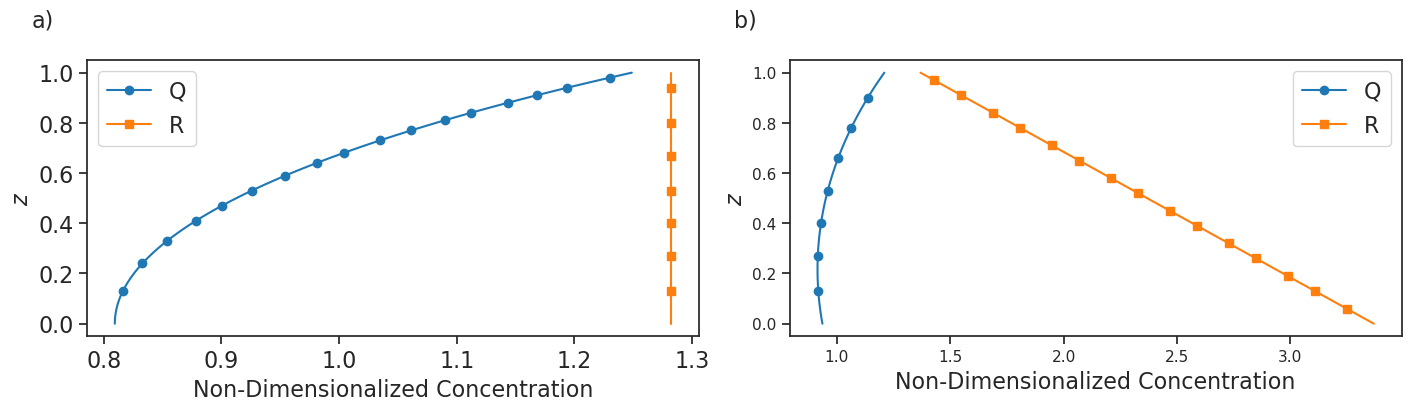}
\caption{\label{fig:steady_state}(a) Unpatterned steady state concentration profiles for the full reaction on the upper boundary. (b) Unpatterned steady states for the separated reaction. In both cases, $a=0.2$, $b=2$, and $\eta_Q=\gamma=D=1$.}
\end{figure}

We present two explicit examples in a two-dimensional domain that correspond to the setup in Figure \ref{fig:schematic_bbr} with nonlinear reactions for both species on the upper boundary. In both cases, we assume $\eta_R = 0$ and thus $D_c(k)$ is determined by \eqref{eq:crit_D_sh}. The reaction occurring on the upper plate is based on the Schnakenberg--Selkov \cite{schnakenberg, selkov} reaction system so that the rates of flux (with physical dimensions of amount of substance per unit length per unit time) are modeled by:
\begin{subequations}\label{eq:dim_schnakenberg_selkov}
\begin{align}
    f(Q,R) &= \alpha-k_{-1}Q+k_2Q^2R,\\
    g(Q,R) &= \beta-k_2Q^2R.
\end{align}
\end{subequations}
We assume that $\alpha,\, \beta,\, k_{-1}$ and $k_2$ are all positive so there is no confusion about which terms are influxes or out-fluxes. This is non-dimensionalized by taking
\begin{equation}
    \begin{IEEEeqnarraybox}[][c]{rClCrClCrCl}
        Q &=& \sqrt{\frac{k_{-1}}{k_2}} \hat{Q}, &\quad& R &=& \sqrt{\frac{k_{-1}}{k_2}}\hat{R}, &\quad& \gamma &=& \frac{H k_{-1}}{d_Q},\\
        \hat{\eta}_Q &=& \frac{H^2}{d_Q}\eta_Q, &\quad& a &=& \sqrt{\frac{k_{2}}{k^3_{-1}}}\alpha, &\quad& b &=& \sqrt{\frac{k_{2}}{k_{-1}^3}}\beta.
    \end{IEEEeqnarraybox}
\end{equation}
Dropping the hats on the non-dimensionalized variables, \eqref{eq:dim_schnakenberg_selkov} becomes,
\begin{subequations}
\begin{align}
    f(Q,R) &= \gamma(a-Q+Q^2R),\\
    g(Q,R) &= \gamma(b-Q^2R).
\end{align}
\end{subequations}
Placing the entire reaction on the upper boundary at $z=1$ yields the first example, and serves as an example of a full reaction presented in Figure \ref{fig:schematic_bbr}:
\begin{equation}\label{eq:sb_nondim}
    \begin{IEEEeqnarraybox}[][c]{rClCrCl}
    f_1(Q,R) &=& a-Q+Q^2R, &\qquad& g_1(Q,R) &=& b-Q^2R,\\
    f_0(Q,R) &=& 0, &\qquad& g_0(Q,R) &=& 0.
  \end{IEEEeqnarraybox}
\end{equation}
Note that these functions are scaled by the parameter $\gamma$ for the rates of reaction, as defined in \eqref{eq:general_bc_bbr}. The steady state is given by:
\begin{subequations}\label{eq:sb_stdy_states}
    \begin{align}
        Q_0(z) &= \gamma\frac{(a+b)}{\Gamma_Q\sinh(\Gamma_Q)+\gamma\cosh(\Gamma_Q)}\cosh(\Gamma_Q z),\\
        R_0(z) &= \frac{b}{Q_0(1)^2},
    \end{align}
\end{subequations}
where $\Gamma_Q = \sqrt{\eta_Q}$.

In the second example, we will place the constant $(a,b)$ terms on the bottom boundary which gives an example of a separated reaction:
\begin{equation}\label{eq:cf_nondim}
  \begin{IEEEeqnarraybox}[][c]{rClCrCl}
    f_1(Q,R) &=& -Q+Q^2R, &\qquad& g_1(Q,R) &=& -Q^2R,\\
    f_0(Q,R) &=& a, &\qquad& g_0(Q,R) &=& b,
  \end{IEEEeqnarraybox}
\end{equation}
The steady state is
\begin{subequations}\label{eq:cf_stdy_states}
   \begin{align}
        Q_0(z) &= \gamma\frac{\Gamma_{Q} a \cosh{\left(\Gamma_{Q} \left(z - 1\right) \right)} + \Gamma_{Q} b \cosh{\left(\Gamma_{Q} z \right)} - a \gamma \sinh{\left(\Gamma_{Q} \left(z - 1\right) \right)}}{\Gamma_{Q} \left(\Gamma_{Q} \sinh{\left(\Gamma_{Q} \right)} + \gamma \cosh{\left(\Gamma_{Q} \right)}\right)},\\
        R_0(z) &= \frac{b \left(\Gamma_{Q} \sinh{\left(\Gamma_{Q} \right)} + \gamma \cosh{\left(\Gamma_{Q} \right)}\right)^{2}}{\gamma^2 \left(a + b \cosh{\left(\Gamma_{Q} \right)}\right)^{2}} + \frac{\gamma b}{D}(1-z).
    \end{align}  
\end{subequations}
The steady state concentration profiles for the two examples are presented in Figure \ref{fig:steady_state} for fixed parameter values. There are some key differences between the two examples.  For the separated reaction, we find that $Q_0(z)$ is minimized within the bulk, while for the full reaction, it is minimized at the non-reactive boundary. Additionally, $R_0(z)$ is constant in the full reaction and linear in the separated reaction.

The partial derivatives appearing in the linearized boundary conditions are the same for both examples, namely,
\begin{equation}\label{eq:ex_bd_linearized}
  \begin{IEEEeqnarraybox}[][c]{rClCrCl}
    f_{1,Q} &=& 2Q_0(1)R_0(1)-1, &\qquad& f_{1,R} &=& Q_0(1)^2,\\
    g_{1,Q} &=& -2Q_0(1)R_0(1), &\qquad& g_{1,R} &=& -Q_0^2(1).
  \end{IEEEeqnarraybox}
\end{equation}
Note that, from~\eqref{eq:sb_stdy_states} and~\eqref{eq:cf_stdy_states}, $Q_0(1)$ and $R_0(1)$ are guaranteed to be positive for $a,b,\gamma > 0$. It follows that $f_{1,R}$ and $g_{1,Q}$ have opposite signs, so real values for $\gamma$ require condition~\eqref{eq:sb_real_decay_condition} to be satisfied.

Substituting \eqref{eq:ex_bd_linearized} into inequality~\eqref{eq:d_conditions_sh}, we find that the second term on the left hand side is negative and hence, the first term must also be negative for the inequality to hold. For the full reaction, substituting in the boundary concentrations from~\eqref{eq:sb_stdy_states} leads to the condition for positive values of $D_c(k)$ that
$b>a$ and
\begin{equation}\label{eq:gamma_1_bound}
    \gamma > \gamma_1(k):=\frac{\Gamma_1\tanh(\Gamma_1)(a+b)-2\Gamma_Q b\tanh(\Gamma_Q)}{b-a}.
\end{equation}

For the separated reaction, we use~\eqref{eq:cf_stdy_states} to obtain the conditions for positive values of $D_c(k)$ that
$b\cosh(\Gamma_Q) > a$ and
\begin{equation}
    \gamma > \gamma_2(k) := \frac{\Gamma_1\tanh(\Gamma_1)(a+b\cosh(\Gamma_Q))-2\Gamma_Qb\sinh(\Gamma_Q)}{b\cosh(\Gamma_Q)-a}.\label{eq:gamma_cond_sr}
\end{equation}

\begin{figure}[t]%
\centering
\includegraphics[width=\textwidth]{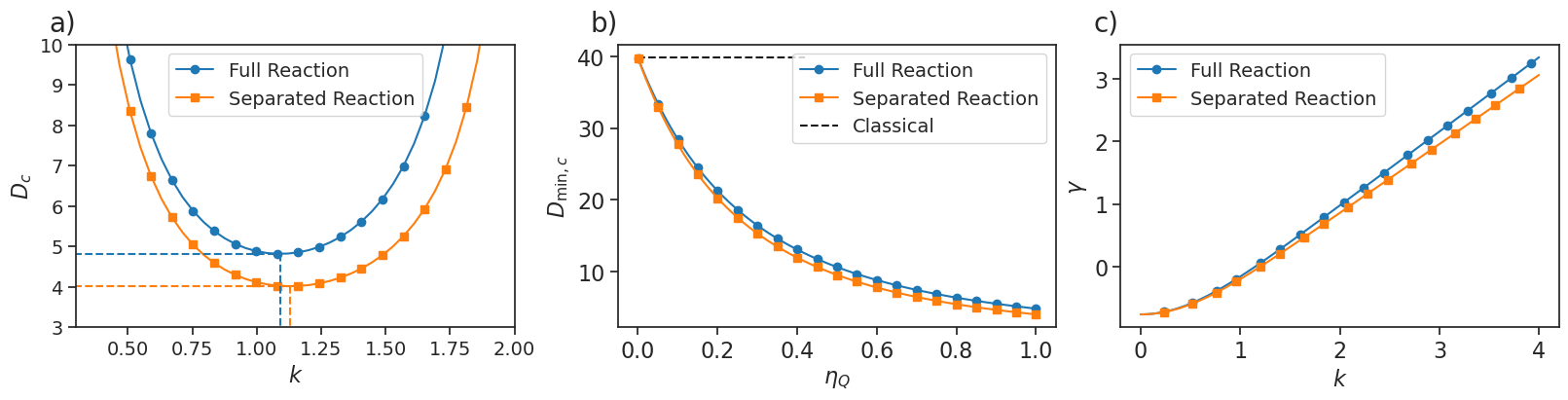}
\caption{\label{fig:crit_D_figures}Comparison of critical parameter values for the full and separated Schnakenberg--Selkov reaction systems. (a) Critical diffusion ratio as a function of wavenumber $k$. (b) Minimum critical diffusion ratio, $D_{\min,c}$, as a function of $\eta_Q$. We also mark the critical diffusion ratio for the classical Turing instability with the Schnakenberg--Selkov reaction system under the same conditions. (c) Global flux thresholds $\gamma_1(k)$ and $\gamma_2(k)$ for the full and separated reactions. In all cases, $(a,b) = (0.2, 2.0)$. When unspecified, $\eta_Q$ and $\gamma$ are set to $1$.}
\end{figure}

Since $\Gamma_1 = \Gamma_Q$ when $k=0$, we have that $\gamma_1(0) < 0$ and hence, $\gamma_1(k) < 0$ for all sufficiently small positive wavenumbers. Similarly, $\gamma_2(k) < 0$ for all sufficiently small positive $k$. Additionally, both $\gamma_1$ and $\gamma_2$ increase monotonically to $+\infty$ with $k$ so that for any fixed $\gamma>0$, there is a critical wavenumber $k_{c,i}>0$ at which $\gamma = \gamma_i(k_{c,i})$, for $i=1,\, 2$. Hence, positive values of $D_c(k)$ exist for all $k$ between 0 and $k_{c,i}$ but not for any wavenumber greater than $k_{c,i}$. In Figure \ref{fig:crit_D_figures} we compare $D_c(k)$ between the two examples. In both cases, after substituting \eqref{eq:ex_bd_linearized} into \eqref{eq:crit_D_sh}, $D_c(k)$ can be expressed as, 
\begin{equation}\label{eq:crit_D_for_examples}
    D_c(k) = \frac{\gamma Q_0(1)^2(\Gamma_1\tanh(\Gamma_1)+\gamma)}{k\tanh(k)(2\gamma Q_0(1)R_0(1) - \gamma - \Gamma_1\tanh(\Gamma_1))}.
\end{equation}
We note that $Q_0(1)$ and $R_0(1)$ depend on $a, b, \gamma$, and $\eta_Q$, but do not depend on $D$ or $k$. By considering the ratio of $D_c(k)$ given by \eqref{eq:crit_D_for_examples} for full and separated reactions over the range of reaction parameter values and wavenumbers where $D_c(k)$ is positive for both, it is possible to show that the critical diffusion ratio for the full reaction is strictly larger than for the separated reaction, as shown in Figure \ref{fig:crit_D_figures}(a) for one parameter set. As a consequence, instabilities can arise with smaller diffusion ratios under the separated reaction compared with the full reaction. Moreover, we show that $k_{c,1} < k_{c,2}$ for fixed model parameters. Suppose that $\gamma = \gamma_1(k_{c,1}) = \gamma_2(k_{c,2})$ and let $\Gamma_1^{(i)}=\sqrt{\eta_Q+k_{c,i}^2}$ denote the value of $\Gamma_1$ corresponding to the threshold wavenumber $k_{c,i}$ for $i=1,2$. Rearranging \eqref{eq:gamma_1_bound} and \eqref{eq:gamma_cond_sr}, we have that
\begin{IEEEeqnarray}{rCl}
    \Gamma_1^{(2)}\tanh(\Gamma_1^{(2)})-\Gamma_1^{(1)}\tanh(\Gamma_1^{(1)})&=&2b\Gamma_Q \sinh(\Gamma_Q) \left(\frac{1}{a+b\cosh(\Gamma_Q)}-\frac{1}{(a+b)\cosh(\Gamma_Q)}\right)\nonumber \\
    && + 2a\gamma\left(\frac{1}{a+b}-\frac{1}{a+b\cosh(\Gamma_Q)}\right),    
\end{IEEEeqnarray}
which is positive if $a,b,\gamma,\Gamma_Q > 0$. Since $\Gamma_1\tanh(\Gamma_1)$ increases monotonically with $k$,  we conclude that $k_{c,2} > k_{c,1}$ and hence, a larger range of wavenumbers can be made unstable using the separated reaction system compared with the full reaction.

\subsubsection{Pattern Forming Regions and Oscillations}

\begin{figure}[ht]%
\centering
\includegraphics[width=\textwidth]{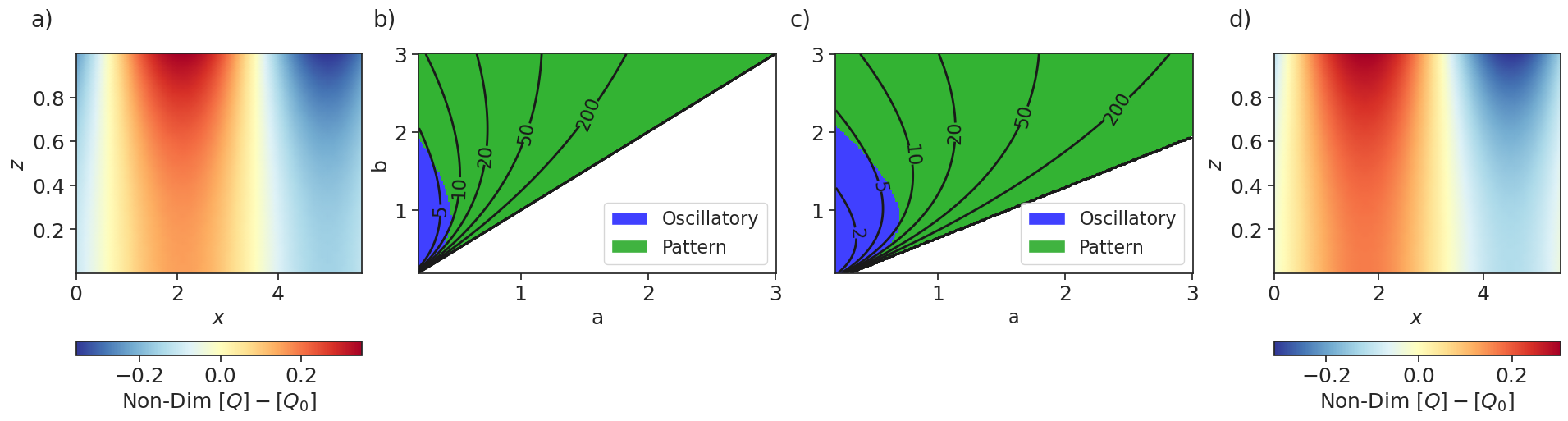}
\caption{\label{fig:pattern_forming_regions} (a) Concentrations relative to unpatterned steady state, taken after the system reaches a steady patterned state for the full Schnakenberg--Selkov reaction with $D=5.1$ and $(a,b) = (0.2, 2.0)$. (b) Pattern forming regions for the full reaction against various $(a,b)$. Contours indicate the minimum critical diffusion ratio for pattern formation. No patterns are expected below the line $a=b$. (c) Pattern forming regions for the separated reaction, with contours for the minimum critical diffusion ratio. No patterns are expected below the line $a=b\cosh(\Gamma_Q)$. (d) Concentrations relative to the unpatterned steady state, taken after the system reaches a steady patterned state for the separated reaction with $D=4.2$ and $(a,b) = (0.2, 2.0)$. We fix the reaction parameters $(\eta_Q, \gamma) = (1,1)$ in all cases.}
\end{figure}

An example of the patterned state emerging when a finite wavenumber becomes unstable in the full reaction system without flow is shown in Figure~\ref{fig:pattern_forming_regions}(a). Since we are interested in using the chemical instabilities to influence convective ones, we focus our attention on parameters $(a,b)$. Varying these parameters can be interpreted as changing the rate of influx of reagents through a selectively permeable boundary, or the rate of production of the reagents from precursors through boundary-bound reactions.

We perform several steps to classify each $(a,b)$ pair in Figure \ref{fig:pattern_forming_regions}(b,c). First, we calculate the critical diffusion and use it to solve for the critical decay for chemical oscillations, see Appendix \ref{ap:osc} for a discussion on chemical oscillations in the model. If the critical decay rate for oscillations is greater than the corresponding reaction parameter, $\eta_{Q,c} > \eta_Q$, it follows that the oscillatory mode is stable at the critical diffusion ratio. However, if $\eta_{Q,c} \leq \eta_Q$ then it is unstable or marginally stable at the critical diffusion. Hence, we classify the regions where $\eta_{Q,c} < \eta_Q$ as oscillatory. We also provide a snapshot of the patterns formed for supercritical diffusion ratios in each example in Figure \ref{fig:pattern_forming_regions}(a,d).

From Figure \ref{fig:pattern_forming_regions}, we note that there are three regions of the $a$--$b$ parameter space where we respectively expect stable unpatterned states, oscillatory behaviour, and steady patterned states to emerge. Thus, we have control over the chemical instability by varying the influxes at the boundaries, $(a,b)$. Interestingly, oscillations are only expected to occur for small values of $a$ and for $b$ within a bounded interval that depends on $a$. Steady patterns can emerge for larger values of $a$ and $b$, provided that the diffusion ratio $D$ is large enough. Below a certain value of $D$, the oscillatory mode is unstable whenever the patterned state is marginally stable so we do not expect steady patterns. Comparing the behaviour of the full and separated reaction systems [Figure~\ref{fig:pattern_forming_regions}(b) and (c), respectively], we observe that for any value of $D$, a larger region of $a$--$b$ parameter space supports oscillations or patterns. Apart from scaling, the regions of parameter space are qualitatively the same for the full and separated systems.

\section{Buoyancy-Driven Instabilities}\label{sec:w_hydrodynamics}

If the Rayleigh number is not negligible, then we cannot omit the momentum equation in \eqref{eq:full_bbr} and must consider the full expression in \eqref{eq:perturbed_equations}. In the absence of a diffusion-driven instability, each reagent can produce convective cells through a mechanism similar to the classical Rayleigh-B\'enard instability and similar to the buoyancy-driven instability described for a linear boundary-bound dissociation reaction~\cite{bdzil_bbr}. In Section \ref{sec:pattern_formation} we discussed limiting cases with constant or linear reactions and highlighted the importance of the nonlinear reaction terms. In light of this, we will focus on the role the diffusive instability plays for the examples presented in Section \ref{sec:exampels_wo_hydrodynamics} and use the case where $D = 1$ to remove the effect of differences in diffusion coefficients, which is a hallmark of traditional Turing instabilities. Recall in Section \ref{sec:pattern_formation} we showed that in the absence of buoyancy forces the system with $D=1$ has no finite wavelength instability.

We again search for marginally stable states ($\lambda=0$) with $\eta_R = 0$ so that \eqref{eq:perturbed_equations} reduces to:
\begin{subequations}\label{eq:ra_equations}
  \begin{align}
    \left(\diff[2]{}{z} - (\eta_Q + k^2)\right) q &= w\diff{Q_0}{z},\\[15pt]
    \left(\diff[2]{}{z} - k^2\right)r &= w\diff{R_0}{z},\\[15pt]
    \left(\diff[2]{}{z} - k^2\right)^2w &= k^2\mathrm{Ra}(\delta^Q q+\delta^R r).\label{eq:w_ra_equations}
  \end{align}
\end{subequations}
If both of the steady-state concentrations are constant, then the reagent concentrations are unaffected by the flow in the linearized system; the linear stability would be identical to that of the system without buoyancy effects. In order for chemo-hydrodynamic interactions to occur at first order, one of $\eta_Q$ or $\eta_R$ must be non-zero, or the model should include a non-trivial in-flux through the non-reactive plate so that these gradients can be non-zero. By construction, both the full and separated Schnakenberg reactions described in \ref{sec:exampels_wo_hydrodynamics} satisfy these conditions. We will refer to a value of $\mathrm{Ra}$ that satisfies \eqref{eq:ra_equations} as a critical Rayleigh number for the wavenumber $k$, denoted $\mathrm{Ra}_c(k)$. We solve for $\mathrm{Ra}_c(k)$ using a shooting method \cite{Stoer2002} applied to \eqref{eq:ra_equations}. The boundary conditions are obtained by calculating \eqref{eq:bbr_bcs} for the examples in \ref{sec:exampels_wo_hydrodynamics}. From our numerical results, we find that the function $\mathrm{Ra}_c(k)$ has at most one local maximum [see Appendix \ref{ap:C}, Figure~\ref{fig:num_v_anal}(a)], which we refer to as the maximum critical Rayleigh number and denote by $\mathrm{Ra}_{\max, c}$, and at most one local minimum, denoted $\mathrm{Ra}_{\min, c}$ and referred to as the minimum critical Rayleigh number. We expect the unpatterned state to become unstable and transition to steady patterns with finite wavenumber as the Rayleigh number increases above the minimum critical Rayleigh number or decreases below the maximum critical Rayleigh number. The existence and values of these minima and maxima depend on parameters such as $\delta^Q$, $\delta^R$, $D$ and $\gamma$. We discuss the performance of the estimates obtained by this shooting method procedure in Appendix \ref{ap:C}.

\begin{figure}[tb]%
\centering
\includegraphics[width=\textwidth]{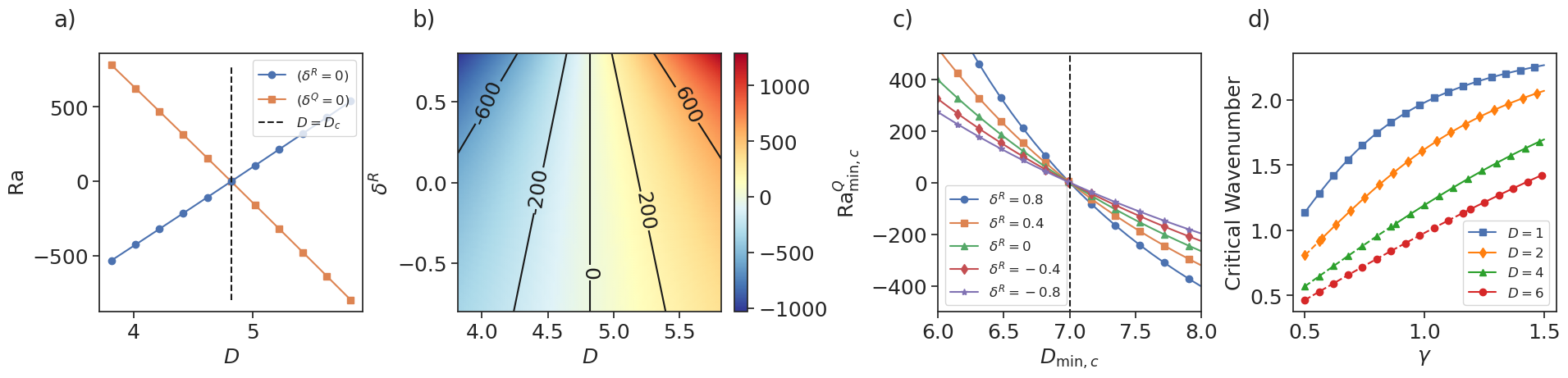}
\caption{\label{fig:q_v_r_driven_flows_buoyancy} (a) Comparison of $\mathrm{Ra}_{\max,c}^Q$ for $Q$-driven flow $(\delta^Q,\ \delta^R)=(1,0)$ and $\mathrm{Ra}_{\min,c}^R$ for $R$-driven flow $(\delta^Q,\ \delta^R)=(0,1)$ for various diffusion ratios. (b) The variation of $\mathrm{Ra}_{\min,c}^Q$ with $D$ and $\delta^R$ for the full reaction system. Reaction parameters are held constant at $a=0.2$, $b=2$, $\eta_Q=1$ and $\gamma=1$. (c) $\mathrm{Ra}_{\min,c}^Q$ as a function of $D_{\min,c}(a,b)$ for various $\delta^R$. We take $(a,b)$ along the line between $(0.268, 2.085)$ and $(0.352, 2.19)$ and $(\gamma,\eta_Q, D)=(1,1,7)$. The dashed line indicates where the system transitions from pattern forming to non-pattern forming without buoyancy forces. (d) Critical wavenumber for convective cells as a function of $\gamma$. We fix $(a,b,\eta_Q) = (0.2, 2, 1)$. }
\end{figure}

In Section \ref{sec:gm} we introduced the notion of $Q$-driven flow with $\delta^Q = 1$. This corresponds to when $Q$ has a larger effect on the buoyancy force at similar concentrations compared to $R$ at similar concentrations. Letting $\delta^R = 0$, we find that there is a maximum but no minimum critical Rayleigh number. The maximum, $\mathrm{Ra}_{\max,c}^Q$, increases with $D$. It is negative, then zero, and positive when the diffusion ratio is respectively below, equal to, and above the minimum critical diffusion ratio, $D_{\min,c} := \min_{k} D_{c}(k)$ [Figure~\ref{fig:q_v_r_driven_flows_buoyancy}(a)]. Note that $D_{\min,c}$ is the diffusion ratio at which the system is marginally stable without flow. When $D < D_{\min,c}$, the unpatterned state is stable without flow but a finite wavenumber can be brought to marginal stability with flow if gravity acts downward (i.e., the Rayleigh number is negative). In this case, coupling with fluid flow serves to destabilize the unpatterned state. Conversely, when $D > D_{\min,c}$, the system without flow is unstable but can be stabilized by flow if gravity acts upwards.

The effect of fluid flow is reversed when $\delta^Q=0, \delta^R=1$; we find that there is a minimum critical Rayleigh number, $\mathrm{Ra}_{\min,c}^R$, that is positive when $D < D_{\min,c}$ and negative when $D > D_{\min,c}$ so flow is destabilizing below the minimum critical diffusion ratio if gravity acts upwards and flow is stabilizing above the minimum critical diffusion ratio if gravity acts downwards. The fact that instabilities can be induced by buoyancy effects when $\delta^Q=0$ and $\delta^R=1$ is interesting because the concentration field $R$, and hence the fluid density, is spatially uniform in the unpatterned steady state. This contrasts with Rayleigh--B\'enard instabilities, in which the instability occurs when the vertical gradient in density at the steady state exceeds a threshold.

The stabilizing and destabilizing effects of $Q$ and $R$ combine when $\delta^Q$ and $\delta^R$ are both nonzero, as shown in Figure~\ref{fig:q_v_r_driven_flows_buoyancy}(b). For instance, given a fixed diffusion ratio $D < D_{\min,c}$, the maximum critical Rayleigh number for $\delta^Q=1$ becomes more negative as $\delta^R$ increases because the stabilizing effect of $R$ (when gravity acts downwards) offsets the destabilizing effect of $Q$.

In Section \ref{sec:pattern_formation}, we discussed how travelling along paths in $(a,b)$ allow us to control the various chemical instabilities. The relationship between $\mathrm{Ra}_c$ and $D_{\min,c}$ makes it possible to control the convective instability not only by changing gradients, but also by varying $D_{\min,c}$ through the parameters $(a,b)$. In Figure \ref{fig:q_v_r_driven_flows_buoyancy}(c), we do precisely this along a line of $(a,b)$ values and plot it for various $\delta^R$ with a fixed $D$. This plot illustrates the relationship between $D_{\min,c}$ and $\mathrm{Ra}$ and demonstrates how $\delta^R$ makes the system more or less sensitive to changes in $D_{\min,c}(a,b)$. This effect could be predicted from Figure \ref{fig:q_v_r_driven_flows_buoyancy}(a,b) since the corresponding $R$-driven flow is stable or unstable depending on whether we are left or right of $D_{\min,c}$.

Whereas for the classical Rayleigh-B\'enard instability the critical wavenumber is fixed, for Turing pattern formation, it can be tuned by the reaction rate. This parameter corresponds to our $\gamma$, which can be interpreted as the enzyme density on the reactive plate. Our model shares this characteristic that the critical wavenumber depends on $\gamma$, as shown in Figure \ref{fig:q_v_r_driven_flows_buoyancy}(d). Similar to classical Turing pattern formation, we find a positive correlation between $\gamma$ and the critical wavenumber.

\begin{figure}[tb]%
\centering
\includegraphics[width=\textwidth]{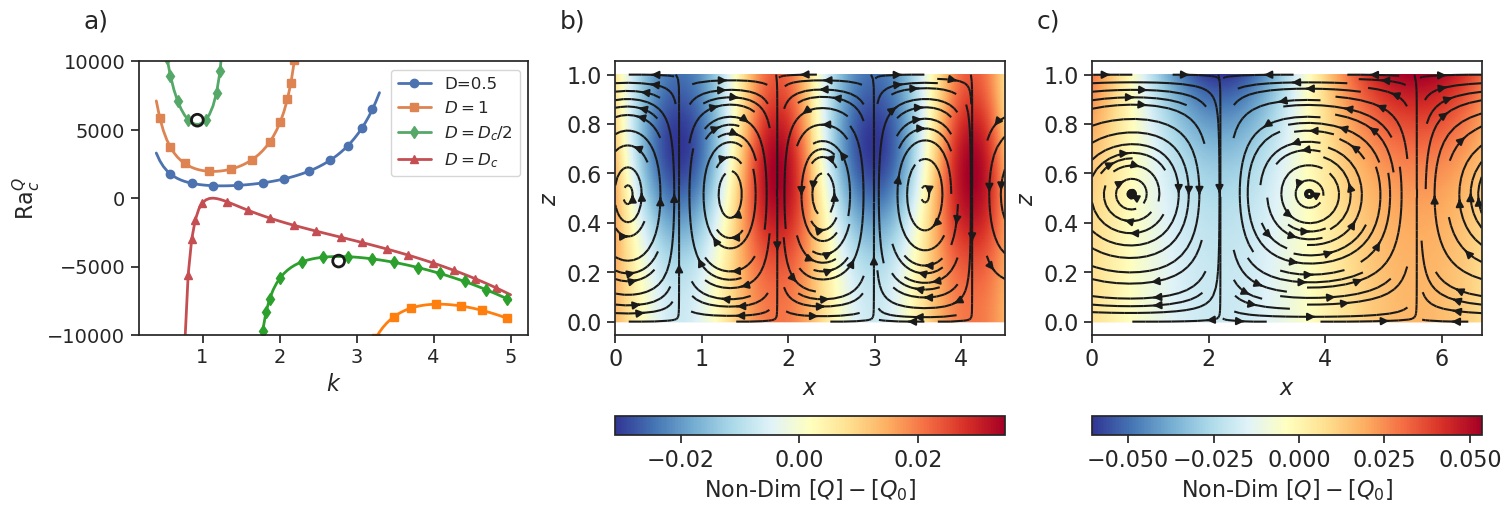}
\caption{\label{fig:neg_pos_comparison}(a) Critical Rayleigh number for the separated reaction with various diffusion ratios, reaction parameters are $a=0.2$, $b=2.0$, $\eta_Q=1$, $\gamma=1$ subject to $Q$-driven flow $(\delta^Q, \delta^R) = (1,0)$. The critical diffusion ratio is $D_c \approx 4.02$. The wavenumbers and Rayleigh numbers in (b) and (c) are marked with open circles. (b) Simulation of unstable system for separated reaction with negative Rayleigh number, $\mathrm{Ra}=-4600$, marked in (a) and $D=D_c/2\approx2.01$. (c) Simulation of unstable system for separated reaction with positive Rayleigh number $\mathrm{Ra}=5700$ marked in (a) and $D=D_c/2\approx2.01$.}
\end{figure}

When we consider the separated reaction, we see other interesting behaviours. Focusing on $Q$-driven flows, in Figure \ref{fig:neg_pos_comparison}(a) we plot the critical Rayleigh number, $\mathrm{Ra}_{c}^Q(k)$, for various wavenumbers and diffusion ratios. We find that it is possible for both minimum and maximum critical Rayleigh numbers to exist, with different associated critical wavenumbers. For example, when $D=D_c/2$, there is a positive minimum critical Rayleigh number and a negative maximum critical Rayleigh number. Between these Rayleigh numbers, the unpatterned state is stable and outside this interval, patterns emerge, illustrated in Figure~\ref{fig:neg_pos_comparison}(b)--(c). As demonstrated, different modes can be made unstable depending on the orientation of the domain relative to the direction of gravity. We suspect that this behaviour occurs because there is a local minimum of $Q$ in the bulk at the steady state [see Figure \ref{fig:steady_state}(b)]. As a result, the concentration gradient changes sign within the domain, allowing it to become unstable regardless of the direction of gravity. We noted above, however, that concentration gradients at the steady state are not a requirement for instabilities.

\section{Conclusions}

We presented boundary-bound reactions and their capacity for chemical and convective pattern formation. We showed that the two types of instabilities are coupled such that buoyancy effects can induce chemical patterns and chemical patterns induce convective flows. For specific cases, we determined necessary conditions for pattern formation in terms of the reaction parameters. We illustrated these results by considering two examples, both based on the Schnakenberg--Selkov reaction system. The critical diffusion ratio was reduced by increasing the bulk-decay $\eta_Q$, similar to results for active-membrane models. We further discussed how the instabilities could be controlled through the constant boundary flux terms $(a,b)$.

Since the chemical instabilities were a sub-case of the buoyancy-driven instabilities, this provided a natural transition for discussing mechanisms for controlling the convective instabilities that form. We also discussed how the parameter $\gamma$, which in practice could be easily tuned through enzyme concentrations or temperature, can be used to change the unstable wavenumber for the convective pattern. Moreover, we observed how the sensitivity to changes in $D_{\min,c}(a,b)$ is influenced by the ratio $\delta^Q/\delta^R$.

Unique to boundary-bound reactions are the nonlinear boundary conditions that influence the buoyancy of the fluid. These nonlinearities were essential for producing behaviour like the buoyancy-driven instability without steady state density gradients. The use of chemical reagents also made it possible to present a system where the formation of convective cells took place independent of the direction of gravity. We presented this as a system which can be used for detecting the orientation of the device. For instance, in Figure \ref{fig:neg_pos_comparison}(b,c) the direction of flow near the peaks and troughs changes depending on the direction of gravity. By measuring the flow field near the boundary as well as the concentration, we can determine which instability is present and as a consequence its orientation. The unstable wavenumbers are also different depending on the direction of gravity. This responsiveness to orientation means it may be possible to design microfluidic devices which change their behaviour by a simple act of turning them upside down, rather than stopping altogether.

We opted to consider reaction parameters that can be thought to physically represent the amount of material pumped into the domain, or the rate of the reaction at the boundary. The latter is often tuned by changing the enzyme density on the surface. In both cases, we demonstrate how one might design systems where the flow fields are controlled via the reaction parameters. Recent work in the design of reaction--diffusion programs provides some freedom in reactions that occur along the boundary. By designing reactions with properties of interest, such as traveling waves or other pattern forming instabilities many other phenomena should be possible. We believe that boundary-bound reactions can facilitate the design of microfluidic devices with novel complex and self-regulating functionality.

\begin{acknowledgments}
We acknowledge the support of the Natural Sciences and Engineering Research Council of Canada (NSERC), [funding reference number RGPIN-2018-04418].

Cette recherche a \'{e}t\'{e} financ\'{e}e par le Conseil de recherches en sciences naturelles et en g\'{e}nie du Canada (CRSNG), [num\'{e}ro de r\'{e}f\'{e}rence RGPIN-2018-04418].
\end{acknowledgments}

\appendix

\section{Reduced Equations for Boundary-Bound Reactions}\label{ap:A}

Substituting Equation \eqref{eq:perturbation} into Equation \eqref{eq:full_bbr} we collect the first order terms:
\begin{subequations}\label{eq:flow_pert_before_reduc}
  \begin{align}
    \diffp{Q_1}{t} &= -\eta_Q Q_1 + \nabla^2 Q_1 - w\diff{Q_0}{z},\\
    \diffp{R_1}{t} &= -\eta_R R_1 + \nabla^2 R_1 - w\diff{Q_0}{z},\\
    \diffp{\mathbf{u}_1}{t} &= -\nabla p_1 + \mathrm{Sc}\nabla^2\mathbf{u}_1+\mathrm{Sc}^2\mathrm{Gr}\left(\delta^QQ_1 + \delta^R R_1\right)\mathbf{e}_z.\label{eq:flow_pert_before_reduc_velocity}
  \end{align}
\end{subequations}
The only term driving fluid flow is the buoyancy force. Therefore, the stability of \eqref{eq:flow_pert_before_reduc} is dependent on the stability of the reduced system containing only the vertical velocity component. Omitting the horizontal flow components, if this reduced system is unstable, then the global system will be as well. To eliminate the pressure term, we consider the divergence of both sides of Equation \eqref{eq:flow_pert_before_reduc_velocity}. By applying the incompressibility condition, we obtain the equations
\begin{equation}
    \nabla^2 p_1 = \mathrm{Sc}^2\mathrm{Gr}\left(\delta^Q\diffp{Q_1}{z}+\delta^R\diffp{R_1}{z}\right).
\end{equation}
Differentiating with respect to $z$, we obtain
\begin{equation}\label{eq:laplacian_pz}
  \nabla^2 \diffp{p_1}{z} = \mathrm{Sc}^2\mathrm{Gr}\left(\delta^Q\diffp[2]{Q_1}{z}+\delta^R\diffp[2]{R_1}{z}\right).
\end{equation}
We now have an equation for the Laplacian of $\diffp{p_1}{z}$ which we can use to eliminate the pressure term. In particular, by taking the Laplacian of the vertical ($z$) component of \eqref{eq:flow_pert_before_reduc_velocity}, we obtain
\begin{equation}
    \diffp{}{t}(\nabla^2 u_{1,z}) = -\nabla^2\diffp{p_{1}}{z} + \mathrm{Sc}\nabla^4 u_{1,z} + \mathrm{Sc}^2\mathrm{Gr}\nabla^2\left(\delta^QQ_1 + \delta^R R_1\right).\\
\end{equation}
Substituting \eqref{eq:laplacian_pz} into this gives
\begin{equation}\label{eq:pressure_removed}
  \diffp{}{t}(\nabla^2 u_{1,z}) = \mathrm{Sc}\nabla^4 u_{1,z} + \mathrm{Sc}^2\mathrm{Gr}\nabla_H^2\left(\delta^QQ_1 + \delta^R R_1\right),
\end{equation}
where $\nabla^2_H = \diffp[2]{}{x}+\diffp[2]{}{y}$ is the horizontal Laplace operator. After substituting the ansatz for the perturbation we can recover \eqref{eq:perturbed_equations}.

\section{On Oscillatory Instabilities}\label{ap:osc}

\begin{figure}[t]%
\centering
\includegraphics[width=0.45\textwidth]{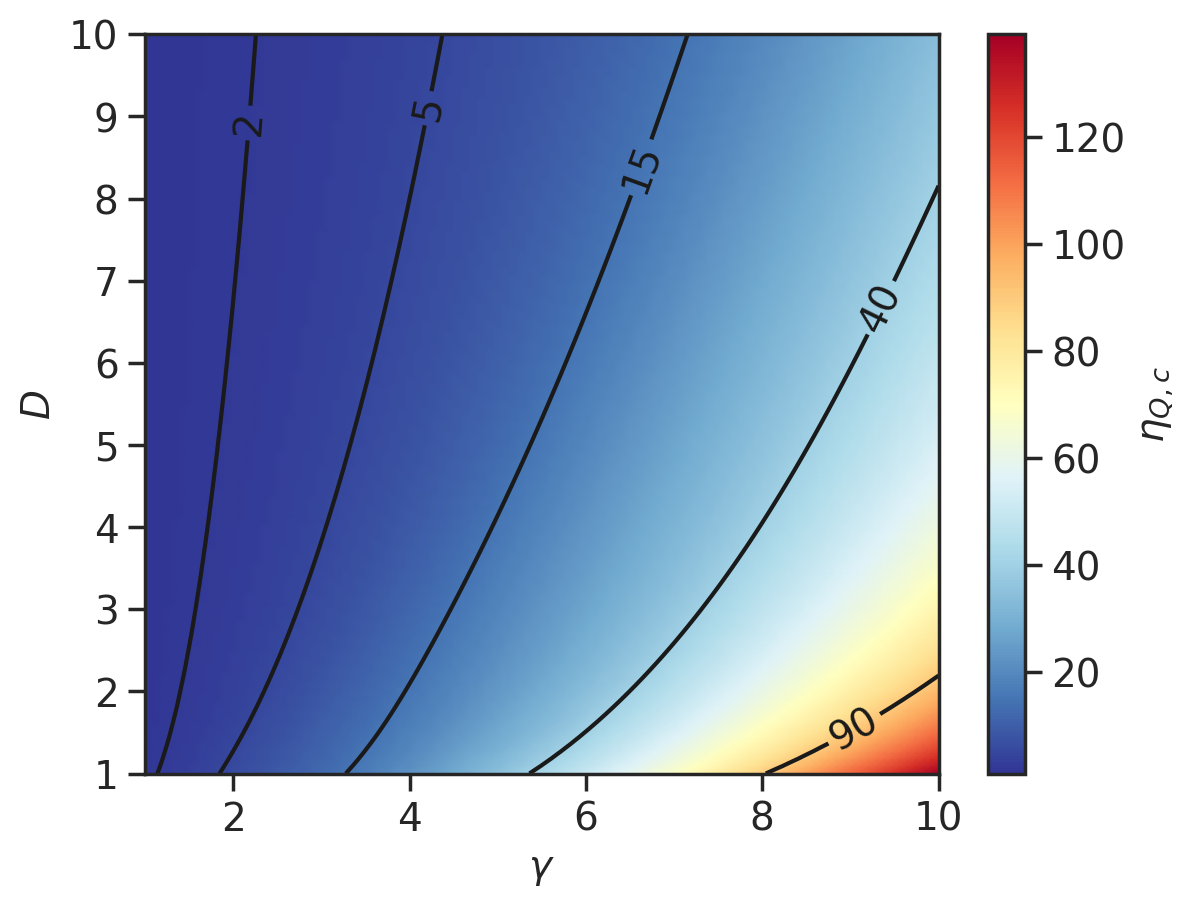}
\caption{\label{fig:crit_eta_osc} Critical values of the decay rate $\eta_Q$ for the full reaction, where long wavelength chemical oscillations become marginally stable as a function of $(\gamma, D)$ with $(a,b)=(0.2, 2)$ fixed.}
\end{figure}

In Turing's analysis \cite{turing}, he was able to determine conditions where chemical oscillations are not possible. Our analysis does not address the stability of the horizontally uniform mode $k=0$. However, long-wavelength chemical oscillations are generally possible and should be expected since they are present in the classical Schnakenberg--Selkov dynamics. To show their existence, we suppose that the height of the fluid chamber is small compared with other length scales and consider the vertically averaged version of \eqref{eq:sb_nondim} and \eqref{eq:cf_nondim}, obtaining for both:
\begin{subequations}
  \begin{align}
    \diffp{\overline{Q}}{t} &= \gamma(a-\overline{Q}+\overline{Q}^2\overline{R}) - \eta_Q \overline{Q} + \nabla^2_H \overline{Q},\\
    \diffp{\overline{R}}{t} &= \gamma(b-\overline{Q}^2\overline{R})+D\nabla^2_H \overline{R}.
  \end{align}
\end{subequations}
The homogeneous steady states are given by
\begin{subequations}
  \begin{align}
    \overline Q_s &= \frac{\gamma(a+b)}{\gamma+\eta_Q},\\
    \overline R_s &= \frac{b(\gamma+\eta_Q)^2}{\gamma^2(a+b)^2}.
  \end{align}
\end{subequations}
Linearizing this equation at the steady state we consider a long-wavelength oscillatory instability. The Jacobian for the system is,
\begin{equation}\label{eq:jacobian_avg}
  J = \begin{bmatrix}
  -(\gamma+\eta_Q)+2\gamma \overline Q_{s}\overline R_s & \gamma \overline Q_{s}^2\\
  -2\gamma \overline Q_{s}\overline R_{s} & -\gamma \overline Q_{s}^2
  \end{bmatrix}.
\end{equation}
A quick calculation shows that $\det(J) > 0$ when $\gamma > 0$. Hence, a Hopf bifurcaton can occur when $\text{Tr}(J)$ passes through zero \cite{Ricard2009}. The trace is
\begin{equation}
  \text{Tr}(J)=\frac{2b(\eta_Q+\gamma)}{(a+b)}-\frac{\gamma^3(a+b)^2}{(\eta_Q+\gamma)^2}-(\eta_Q+\gamma),
\end{equation}
which is zero when $b > a$ and the decay rate $\eta_Q$ takes the critical value
\begin{equation}
  \eta_{Q,c} = \gamma\left(\frac{a+b}{\sqrt[3]{b-a}}-1\right).\label{eq:averaged_crit_eta}
\end{equation}
When $\eta_Q > \eta_{Q,c}$, we find $\text{Tr}(J) > 0$ yielding an unstable focus. This implies the existence of chemical oscillations, which we would also expect to occur in the unaveraged system.

In Figure \ref{fig:crit_eta_osc} we consider the critical decay for chemical oscillations in the full reaction. The estimates are obtained numerically by setting $k=0$ and assuming the form $\lambda = i\omega$ for some real $\omega$ in \eqref{eq:perturbed_equations}, without the fluid momentum equation. We optimize over the $(\eta_Q,\omega)$-plane to find the critical values of $\eta_Q$ as a function of $(\gamma, D)$. The plot shows that $\gamma$ increases the critical value of $\eta_Q$, which we could infer from \eqref{eq:averaged_crit_eta}, and simultaneously shows that $D$ decreases the critical value, something that was not captured by the above vertically averaged analysis.

\section{Comparison of Numerical and Analytical Estimates for Critical Values}\label{ap:C}

We implement a bisection algorithm comparing our estimates to simulations in order to test the critical diffusion and critical Rayleigh number estimates for the full reaction. Each simulation is done using Dedalus \cite{dedalus}, a pseudo-spectral method to solve \eqref{eq:reaction_eq}, with a third order semi-implicit backwards difference method \cite{ruuth1995}. To define initial bounds, we initialize the bisection algorithm with a $1\%$ perturbation above and below the critical values, or $\pm1$ when the critical value is zero; we verify that the unpatterned steady states are unstable and stable, respectively, at the two bracketing points and widen the window as needed. To determine whether patterns have formed, we calculate the amplitude of the pattern at the boundary and use it to implement stopping conditions. The simulation is stopped if the rate of change in the amplitude is sufficiently small relative to its size or the amplitude is smaller than a cutoff, both of which were on the order of $10^{-8}$ or smaller. Additionally, a maximum time was set to end simulations if neither of these conditions were reached. Once the simulation was finished, we compared the final amplitude of the pattern to our cutoff; we classify the system as unstable if it is larger and use this to update the bounds. This means the algorithm generally overestimates the critical parameters slightly, as it cannot distinguish between stable states and patterns with amplitudes below the cutoff. We bisect the interval five times, to obtain an accuracy of $\sim0.156\%$ assuming that the upper and lower bounds were accepted. We report the final midpoint values and corresponding deviations from the estimates obtained analytically or via the shooting method in Tables \ref{tab:crit_diff_num_vs_nm} and \ref{tab:crit_ra_tab} . The similarity between the two critical estimates suggests that our calculations are accurate.

\begin{figure}[t]
  \centering
  \includegraphics[width=\textwidth]{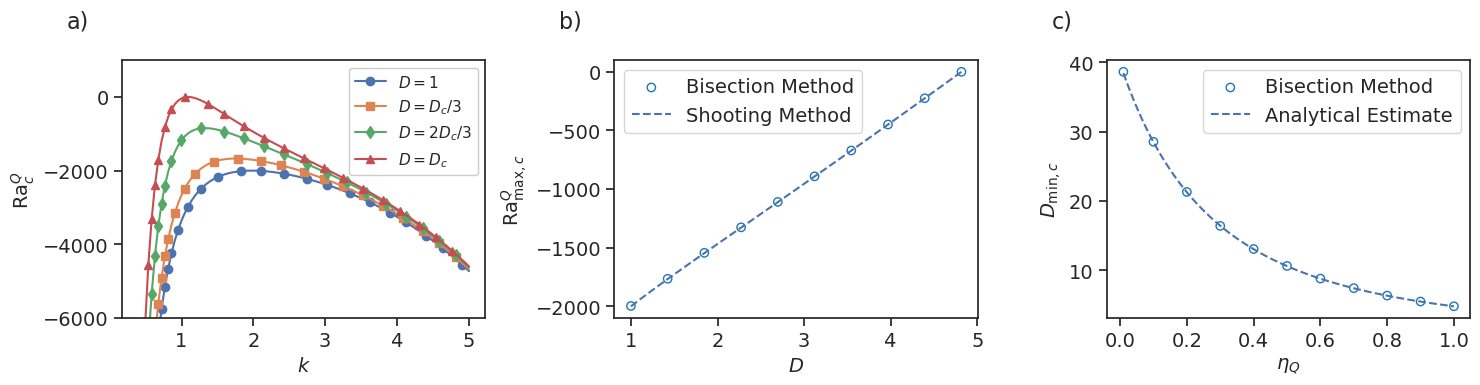}
  \caption{\label{fig:num_v_anal} Critical Rayleigh numbers and diffusion ratios for the full reaction system with $Q$-driven flow $(\delta^Q,\delta^R) = (1,0)$. Except where explicitly varied, reaction parameters are $(a, b, \eta_Q, \gamma) = (0.2, 2, 1, 1)$. (a) Critical Rayleigh number as functions of wavenumber for various diffusion ratios between $1$ and $D_c\approx 4.82$, obtained by a shooting method. (b) Comparison of the maximum critical Rayleigh number obtained by determining the local maxima of the functions in (a) with estimates from the bisection method. (c) Critical diffusion estimates vs.\ $\eta_Q$, where $\mathrm{Ra} = 0$, compared with estimates from the bisection method.}
\end{figure}

\begin{table}[htpb]
  \centering
  \begin{tabular}{c|c|c|c}
      Decay\hspace{1ex} & \hspace{1ex}Diffusion Estimate\hspace{1ex} & \hspace{1ex}Bisection Algorithm\hspace{1ex} & \hspace{1ex}Relative Error\\
      \hline
      0.01 & 38.472 & 38.653 & 0.469\%\\
      0.1 & 28.514 & 28.558 & 0.156\%\\
      0.2 & 21.286 & 21.319 & 0.156\%\\
      0.3 & 16.445 & 16.419 & 0.156\%\\
      0.4 & 13.065 & 13.086 & 0.156\%\\
      0.5 & 10.624 & 10.641 & 0.156\%\\
      0.6 & 8.809 & 8.823 & 0.156\%\\
      0.7 & 7.428 & 7.439 & 0.156\%\\
      0.8 & 6.353 & 6.363 & 0.156\%\\
      0.9 & 5.022 & 5.511 & 0.156\%\\
      1.0 & 4.818 & 4.825 & 0.156\%
  \end{tabular}
  \caption{Comparison of an analytical estimate and the bisection algorithm for the minimum critical diffusion for the full reaction. Reaction parameters used are $(a,b,\gamma) = (0.2, 2, 1)$. Results are rounded to three decimal places.}
  \label{tab:crit_diff_num_vs_nm}
\end{table}

\begin{table}[htpb]
  \centering
  \begin{tabular}{c|c|c|c}
  \hspace{1ex}Diffusion\hspace{1ex} & \hspace{1ex}Estimate\hspace{1ex} & \hspace{1ex}Bisection Algorithm\hspace{1ex} & \hspace{1ex}Absolute Error\hspace{1ex}\\
  \hline
  $1$ & $-1999.659$ & $-1996.534$ & $3.124$\\
  $1.424$ & $-1767.331$ & $-1765.570$ & $2.761$\\
  $1.848$ & $-1546.015$ & $-1543.599$ & $2.415$\\
  $2.273$ & $-1328.441$ & $-1326.365$ & $2.075$\\
  $2.697$ & $-1111.122$ & $-1109.386$ & $1.736$\\
  $3.121$ & $-892.542$ & $-891.148$ & $1.394$\\
  $3.545$ & $-672.156$ & $-671.106$ & $1.050$\\
  $3.969$ & $-449.864$ & $-449.162$ & $0.702$\\
  $4.393$ & $-225.758$ & $-225.406$ & $0.352$\\
  $4.818$ & $0$ & $0.044$ & $0.044$
  \end{tabular}
  \caption{\label{tab:crit_ra_tab} Comparison of the shooting method estimate and the bisection algorithm for $\mathrm{Ra}_{\max,c}^Q$ for the full reaction with $Q$-driven flow with various diffusion ratios $D$ and fixed reaction parameters $(a,b,\eta_Q,\gamma) = (0.2, 2.0, 1, 1)$. Results are rounded to three decimal places.}
\end{table}


\newpage

\bibliography{bib.bib}

\end{document}